\documentstyle[prl,aps,twocolumn,epsfig,floats]{revtex}

\newcommand{\ket}[1]{\, | #1 \rangle}
\newcommand{\mean}[1]{\langle #1 \rangle}
 
\newcommand{\om}{\omega}
\newcommand{\Om}{\Omega}
\newcommand{\La}{\Lambda}
\newcommand{\ga}{\gamma}
\newcommand{\eps}{\epsilon}
\newcommand{\De}{\Delta}
\newcommand{\de}{\delta}
\newcommand{\lra}{\leftrightarrow}

\begin{document}

\draft

\twocolumn[\hsize\textwidth\columnwidth\hsize\csname@twocolumnfalse\endcsname

\title{Symmetric photon-photon coupling by atoms with Zeeman-split sublevels}

\author{David Petrosyan and Gershon Kurizki}

\address{Department of Chemical Physics, Weizmann Institute of Science, 
Rehovot 76100, Israel}

\date{\today}

\maketitle

\begin{abstract}
We propose a simple scheme for highly efficient nonlinear interaction 
between two weak optical fields. The scheme is based on the attainment 
of electromagnetically induced transparency {\it simultaneously for both 
fields} via transitions between magnetically split $F=1$ atomic sublevels,
in the presence of two driving fields. Thereby, equal slow group velocities 
and symmetric cross-coupling of the weak fields over long distances are 
achieved. By simply tuning the fields, this scheme can either yield giant 
cross-phase modulation or ultrasensitive two-photon switching.
\end{abstract}

]

The weakness of optical nonlinearities in conventional media precludes the 
effective interaction of extremely feeble fields containing few photons only
\cite{nonlin}. This weakness sets a limit on the performance of ultrasensitive
photonic elements (switches and couplers), as well as on nonclassical
(``squeezing'') effects. It is also the main impediment towards constructing 
quantum logic gates, quantum teleportation and cryptography schemes operating 
at the few-photon level \cite{qinf}. The weakness of 
optical nonlinearities can be compensated by photon confinement in a high-Q 
cavity \cite{cavity}. A promising avenue has been opened by studies of enhanced
nonlinear coupling via electromagnetically induced transparency (EIT) in 
atomic vapors in the presence of classical driving fields, which induce 
coherence between atomic levels \cite{eit,arimondo}. These studies have 
predicted the ability to achieve an appreciable nonlinear phase shift of 
extremely weak optical fields \cite{imam} or a two-photon switch 
\cite{harris,zhu}, using the driven $N$-configuration of atomic levels. 
The main hindrance of such schemes is the {\it mismatch} between the group 
velocities of the field that is subject to EIT and its nearly-free propagating 
partner, which severely limits their effective interaction length 
\cite{harris_d}. 

In the present paper we propose a scheme that can remove this bottleneck, by
basically modifying the nonlinear interaction of weak optical fields: in 
contrast to all currently known schemes, it affects both fields in a 
completely {\it symmetric} fashion, rendering their group velocities equal. 
It thereby allows their cross-coupling over very long distances and brings it 
to its ultimate limit of efficiency. This scheme relies solely on an 
{\it intraatomic process} that causes simultaneous EIT for both fields 
interacting with magnetically (Zeeman-) split sublevels in the presence of 
two driving fields. Remarkably, by simply tuning the fields, this scheme can 
either yield giant cross-phase modulation or ultrasensitive two-photon 
switching in vapor \cite{lukin}. It may therefore substantially advance the
optical processing and communication of quantum information in conventional 
media, without resorting to photonic crystals \cite{pc} or resonators 
\cite{tomas}.


\begin{figure}[t]
\centerline{\psfig{figure=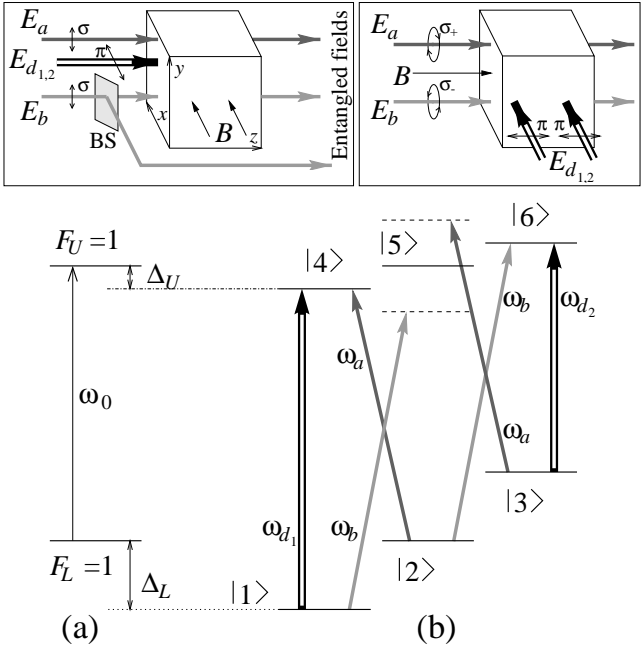,width=8.5cm}}
\vspace{0.2cm}
\caption{Left Inset: Copropagating weak $\sigma$-polarized optical fields 
$E_{a,b}$, and strong $\pi$-polarized driving fields $E_{d_{1,2}}$ pass 
through the atomic vapor cell which is placed in a transverse magnetic 
field $B$. The beam-splitter BS splits the $E_b$ field into two components, 
one of which passes through the cell and interacts with the $E_a$ field. 
The $E_a$ and $E_b$ fields at the output are then entangled.
Right Inset: Perpendicular arrangement of the weak and driving fields
suitable for a cold atomic gas.
(a) Two degenerate atomic levels with angular momenta $F_L=F_U=1$, 
are separated by the unperturbed energy difference $\hbar \om_0$.  
(b) Magnetic field $B$ splits each level into three Zeeman components with 
relative shifts $\De_L$ and $\De_U$. Two $\La$ systems, 
$\La_a \equiv \ket{2} \lra \ket{4} \lra \ket{1}$ and
$\La_b \equiv \ket{2} \lra \ket{6} \lra \ket{3}$, sharing ground state 
$\ket{2}$, are formed in the presence of $E_{d_{1,2}}$($\om_{d_{1,2}}$) and 
give rise to EIT for the $E_a$($\om_a$) and $E_b$($\om_b$) fields, 
respectively. Transitions $\ket{1} \to \ket{5}$ and 
$\ket{3} \to \ket{5}$ serve to cross-couple $\La_a$ and $\La_b$ with the 
$\om_b$ and $\om_a$ photons, respectively.}
\label{system}
\end{figure}

We first outline the proposed setup. Two weak 
optical fields $E_a$ and $E_b$, having $\sigma$ polarization along the $y$ 
axis, propagate in the atomic medium along the $z$ axis. 
Two $\pi$-($x$-)polarized CW driving fields $E_{d_{1,2}}$ and a static 
$x$-oriented magnetic field $B$ are applied (Fig. \ref{system}-Left Inset). 
In the absence of these fields, the lower ($L$) and upper 
($U$) atomic levels, having equal total angular momenta $F_L=F_U=1$ 
(e.g, the $D1$ line in $^{23}$Na or $^{87}$Rb), are separated by the 
frequency $\om_0$ [Fig. \ref{system}(a)]. The magnetic field $B$ splits 
the lower and upper triplets of the atom into three components each, 
labeled $\ket{1}$, $\ket{2}$, $\ket{3}$ and $\ket{4}$, $\ket{5}$, $\ket{6}$, 
respectively [Fig. \ref{system}(b)]. By appropriately tuning the fields, as 
detailed below,  two symmetrically cross-coupled $\La$ configurations, 
$\La_a \equiv \ket{2} \lra \ket{4} \lra \ket{1}$ and 
$\La_b \equiv \ket{2} \lra \ket{6} \lra \ket{3}$, are realized and give rise 
to EIT for the two weak fields simultaneously.

Let us specify the transitions and fields involved. The Zeeman 
shift of the sub-levels in the lower and upper level is given by
$\De_{L,U} = (\mu_B/\hbar) M_{L,U} g_{L,U} B$, where $\mu_B$ is the Bohr 
magneton, $g_{L,U}$ is the gyromagnetic factor of the corresponding atomic 
level and $M_{L,U}$ is the magnetic quantum number of the corresponding state.
The $\sigma$-polarized $E_a$ and $E_b$ fields act only on the transitions 
with $\De M = \pm 1$, while the $\pi$-polarized $E_{d_{1,2}}$ fields couple 
the states with $\De M = 0$ ($M \neq 0$).
The frequencies of the CW driving fields $E_{d_{1,2}}$ are resonant with the 
atomic transitions $\ket{1} \to \ket{4}$ and $\ket{3} \to \ket{6}$, 
respectively: $\om_{d_{1,2}} = \om_0 \pm \De_D$, where $\De_D = \De_L - \De_U$.
The $E_{d_1}$ and $E_{d_2}$ fields, having the same Rabi frequency $\Om_d$,
act also on the transitions $\ket{3} \to \ket{6}$ and $\ket{1} \to \ket{4}$, 
with the detunings $\mp 2 \De_D$, respectively. If  $\De_D \gg |\Om_d|$, 
this off-resonant coupling merely induces the ac Stark shifts 
$\mp |\Om_d|^2/2\De_D$ of the states $\ket{1}$ and $\ket{3}$, which can be 
incorporated into the energy of the atomic state. This reasoning is valid, 
provided the detuning $\De_D$ is larger than the Doppler width of the atomic 
resonance $k\overline{v}$, where $k =\om_0 /c$ and 
$\overline{v} = \sqrt{3k_BT/m}$ is the mean thermal velocity of the atoms.  
The frequencies of the weak fields $E_{a,b}$ are chosen to be at Raman 
resonance with the corresponding two-photon transitions $\ket{2} \to \ket{1}$ 
and $\ket{2} \to \ket{3}$: $\om_{a,b} = \om_0 \mp \De_U +\de_{a,b}$, where 
$\de_{a,b}$ denotes a small detuning from the corresponding two-photon 
transition. 

The cross-coupling of the weak fields is achieved via the $E_b$-induced 
transition $\ket{1} \to \ket{5}$ and the $E_a$-induced transition 
$\ket{3} \to \ket{5}$. The fields are detuned from the corresponding 
transitions by the amounts $\pm \De_D$. The realization of the 
{\it dispersive cross-coupling} requires that $\De_D \gg |\Om_{a,b}|$, where 
$\Om_{a,b}$ is the Rabi frequency of the corresponding field. The undesirable
atom-field couplings via the transitions $\ket{1} \to \ket{5}$ and 
$\ket{2} \to \ket{6}$ for $E_a$ and via the transitions $\ket{3} \to \ket{5}$ 
and $\ket{2} \to \ket{4}$ for $E_b$, can then be neglected, due to the large 
detunings $\pm (\De_L + \De_U)$ and $\pm 2 \De_U$, respectively. 

In an alternative setup, one can get rid of these couplings using the 
(circularly) $\sigma_+$ polarized $E_a$ and $\sigma_-$ polarized $E_b$ fields 
propagating along the magnetic field lines (Fig. \ref{system}-Right Inset). 
Then the $E_a$ field will act only on the transitions with $\De M = +1$, that 
is $\ket{2} \to \ket{4}$ and $\ket{3} \to \ket{5}$, while the $E_b$ field will 
act on $\De M = -1$ transitions, that is $\ket{2} \to \ket{6}$ and 
$\ket{1} \to \ket{5}$. However, in order to cancel the Doppler broadening 
of the EIT resonances in this setup, one would have to work with cold atoms 
($T \leq 0.5$ $\mu$K). By contrast, in the collinear Doppler-free geometry of 
Fig. \ref{system}-Left Inset, the EIT resonances can be very sharp at 
considerably higher temperatures. 

Under optimal conditions for cross coupling in both setups discussed above
(Fig. \ref{system}-Left and Right Insets), we obtain the following set of 
coupled equations for the slowly varying probability amplitudes $A_j$ of the 
six atomic states:
\begin{mathletters}
\label{ampls}
\begin{eqnarray}
\partial_t A_1 &=& i \de_a A_1 - i \Om_d^* A_4 + 
i \Om_b^* e^{-i \De_D t} A_5 , \\
\partial_t A_2 &=& - i \Om_a^* A_4 + i \Om_b^* A_6 , \\
\partial_t A_3 &=& i \de_b A_3 - i \Om_a^* e^{i \De_D t} A_5 
+ i \Om_d^* A_6 , \\
\partial_t A_4 &=& -[\ga - i (\de_a - kv)] A_4 
\nonumber \\ & &
- i \Om_d A_1 - i \Om_a A_2 , \\
\partial_t A_5 &=& -[\ga - i (\de_b +\de_a - kv)] A_5 
\nonumber \\ & &
+ i \Om_b e^{i \De_D t} A_1 - i \Om_a e^{-i \De_D t} A_3 , \\
\partial_t A_6 &=& -[\ga - i (\de_b - kv)] A_6 
\nonumber \\ & &
+ i \Om_b A_2 + i \Om_d A_3 ,  
\end{eqnarray}
\end{mathletters}
where $\ga$ is the relaxation rate of states $\ket{4}$, $\ket{5}$ and 
$\ket{6}$, which we include here {\it phenomenologically} 
\cite{harris,harris_d}. In deriving these 
equations, we have replaced the wave numbers $k_j$ ($j=a,b,d_1,d_2$) by 
$k$ and, consistently with the discussion above, have assumed that the ac 
Stark shifts of the states $\ket{1}$ and $\ket{2}$ due to the off-resonance 
interaction with the $E_{d_2}$ and $E_{d_1}$ fields are incorporated in 
$\De_D$. Obviously, in the cold-atom setup (Fig. \ref{system}-Right Inset)
the velocity of atoms $v$ is zero. In Eqs. (\ref{ampls}), the sign of each 
term containing a Rabi frequency is determined by the Clebsch-Gordan 
coefficient for the corresponding atomic transition.

We neglect the depletion of the CW driving fields, for reasons explained below.
In the perturbative solution of Eqs. (\ref{ampls}), we shall assume that 
the relaxation time of the excited atomic states is short compared to the 
duration $\tau$ of the pulses $E_{a,b}$: $\tau \gg \ga^{-1}$. Therefore
the slowly varying envelope approximation is well justified, since the 
pulse amplitudes change very little during an optical cycle. Then the 
evolution of the two weak fields $E_j$ ($j=a,b$) is governed by the 
following propagation equations
\begin{equation}
[\partial_z + v_g^{-1} \partial_t] E_j = 
i \alpha_j E_j \; , \label{maxw}
\end{equation}
yielding $E_j(z,t) = E_j(0,t-z/v_g) \exp \left(i \int_0^z \alpha_j dz \right)$.
Here the macroscopic complex polarizabilities $\alpha_j$, given by
\begin{mathletters}
\label{alphas}
\begin{eqnarray}
\alpha_a &=& \frac{\alpha_0  \ga}{\Om_a}\mean{A_2^* A_4 + 
A_3^* A_5 e^{i \De_D t}}_T , \\
\alpha_b &=& \frac{\alpha_0  \ga}{\Om_b}\mean{A_2^* A_6 + 
A_1^* A_5 e^{-i \De_D t}}_T , 
\end{eqnarray}
\end{mathletters}
are proportional to the linear resonant absorption coefficient
$\alpha_0 = |\mu|^2 \om_0 N/(2 \eps_0 c \hbar \ga) \equiv \sigma_0 N$, 
where $\sigma_0$ is the resonant absorption cross-section and $N$ is the 
density of atoms. In Eqs. (\ref{alphas}), the averaging $\mean{\ldots}_T$ is 
performed over the atomic thermal motion, which obeys the Maxwellian 
distribution $W(v) = (u \sqrt{\pi})^{-1} \exp(-v^2/u^2)$.

The crucial point about Eq. (\ref{maxw}) is that, due to the {\it symmetry} 
of the system with respect to the two fields $E_a$ and $E_b$, their 
group velocities 
$v_g = [ 1/c + \partial \text{Re}(\alpha_j)/\partial \de_j]^{-1}$ are 
{\it equal}, as can be verified from Eqs. (\ref{alphas}) and (\ref{ampls}). 

We assume that initially the driving fields optically pump all the atoms into 
the energy state $\ket{2}$. In the weak field limit (much less than one 
photon per atom), the Rabi frequencies of the two interacting fields satisfy
the condition $|\Om_{a,b}| \ll \ga,|\Om_d|$. Then, during propagation, nearly 
all of the atomic population remains in $\ket{2}$ ($A_2 \simeq 1$). This 
justifies the neglect of the driving-field depletion (assumed above) after the 
optical pumping is completed. Under these conditions, Eqs. (\ref{ampls}) can 
be solved using fourth-order perturbation theory. Then, at the Raman resonance
for both fields, $\de_a = \de_b = 0$, the polarizabilities are given by
\begin{equation}
\alpha_{a,b} = \frac{2 i \alpha_0 \ga |\Om_{b,a}|^2}
{(\ga \pm i \De_D) |\Om_d|^2} , \label{alpha_sltn}
\end{equation}
and the common group velocity of $E_a$ and $E_b$ becomes 
$v_g \simeq |\Om_d|^2/(\alpha_0 \ga) \ll c$. Since the Doppler width 
$k\overline{v}$ is assumed to be smaller than the detuning $\De_D$, 
the averaging over the atomic thermal motion has practically no bearing on 
Eq. (\ref{alpha_sltn}).

The real part of the complex polarizability $\alpha_j$ is responsible for the 
phase shift $\phi_j$ of the corresponding field, 
$\phi_j(z) \simeq \text{Re}(\alpha_j) z$, while the probability of the 
absorption $p_j$ of the field depends on the imaginary part of $\alpha_j$, 
$p_j(z) \simeq 1 - \exp[-2 \text{Im}(\alpha_j) z]$. From Eq. (\ref{alpha_sltn})
we see that if one of the fields propagates alone in the medium, its 
absorption and phase shift are zero \cite{comment}. By contrast, if both 
fields are present, each of them induces an absorption and a phase shift on 
the other. One can verify from Eqs. (\ref{alpha_sltn}) that  
$\text{Re} (\alpha) /\text{Im} (\alpha) = \pm \De_D / \ga$, 
i.e., for $\De_D \gg \ga$, the {\it phase shift is the dominant process} and 
absorption can safely be neglected. This can be realized taking advantage of
the common {\it anomalous Zeeman effect}, which corresponds to $g_L \neq g_U$ 
and thus $\De_D \neq 0$, as, e.g., in $^{23}$Na or $^{87}$Rb atoms. 
Then we obtain
\begin{mathletters}
\label{Re_Im_app}
\begin{eqnarray}
\text{Re} (\alpha_{a,b}) & \simeq & \pm \frac{2\alpha_0 \ga |\Om_{b,a}|^2}
{ \De_D |\Om_d|^2} , \label{Re_app} \\
\text{Im} (\alpha_{a,b}) & \simeq & \frac{2\alpha_0 \ga^2 |\Om_{b,a}|^2}
{\De_D^2 |\Om_d|^2} .  
\end{eqnarray}
\end{mathletters}

In the discussion pertaining to the collinear Doppler-free setup 
(Fig. \ref{system}-Left Inset), we have neglected the interaction of the 
$E_a$ and $E_b$ fields with the atom via the transitions $\ket{1} \to \ket{5}$ 
and $\ket{3} \to \ket{5}$, respectively, due to the large detunings 
$\pm (\De_L + \De_U)$. This simplification leads to the neglect of
the self-phase modulation of the weak optical fields, given by
\begin{equation}
\text{Re} (\alpha_{a,b}) \simeq  \pm \frac{\alpha_0 \ga |\Om_{a,b}|^2}
{ (\De_L + \De_U) |\Om_d|^2},
\end{equation}
which is a weaker effect than the cross-phase modulation of Eq. (\ref{Re_app}).
Furthermore, this self-phase modulation does not depend on the presence of 
the other field, which allows us to separate it from the cross-phase 
modulation. For some applications, however, such as generation of optical 
solitons and phase conjugation \cite{nonlin}, this self-phase modulation may 
be important and interesting in its own right. We stress that the self-phase 
modulation is absent in the setup of Fig. \ref{system}-Right Inset. 

As a concrete example of cross-phase modulation, consider the vapor of 
$^{87}$Rb atoms, where the pertinent lower and upper levels are 
$5S_{1/2},F_L=1$ and $5P_{1/2},F_U=1$, with gyromagnetic factors 
$g_L = -\frac{1}{2}$ and $g_U = -\frac{1}{6}$, respectively, and the 
transition frequency is $\om_0 \simeq 2 \pi \, 3.775 \times 10^{14}$ rad/s. 
Let us choose $N = 10^{14}$ cm$^{-3}$, $|\Om_d| = 5 \times 10^6$ rad/s, and 
$\De_D = 70 \ga$, corresponding to $B \simeq 430$ G, 
$\De_L \simeq 2 \pi \, 3 \times 10^{8}$ rad/s and 
$\De_U \simeq 2 \pi \times 10^{8}$ rad/s, which are smaller than the hyperfine 
splittings of the lower and upper atomic levels, $6.8 \times 10^9$ s$^{-1}$ 
and $8.1 \times 10^8$ s$^{-1}$, respectively. Yet the Doppler width 
$k\overline{v}$ of the atomic transitions should be at least few times 
smaller than $\De_D$. For the chosen magnetic field $B$, this sets the upper 
limit on the temperature of the atomic gas, which for the parameters listed 
above is $T \leq 10$ K. Stronger magnetic fields would allow 
for higher temperatures, at the expense of longer propagation distance or 
higher density. For the present values, two focused {\it single-photon beams} 
$E_a$ and $E_b$ (beam cross-section $\sigma_{a,b} \simeq 10^{-8}$ cm$^{2}$) 
of a $\mu$s duration can induce a mutual phase shift of the order of $\pi$ 
over a distance of $\sim 3.8$ cm, while their absorption probability remains 
close to zero, $p_j < 0.1$. In the setup with cold atomic gas 
(Fig. \ref{system}-Right Inset) we obtain the same phase shift $\pi$ and 
absorption over a distance of propagation of $\sim 1$ cm \cite{dip_mom}, 
corresponding to the interaction of the fields with $\sim 10^{6}$ atoms.


Next we consider the cross-absorption scheme. The atomic level configuration 
is the same as in Fig. \ref{system}(b), but the frequencies of all the fields 
are lowered by $\De_D$, i.e., $\om_{d_1} = \om_0$, 
$\om_{d_2} = \om_0 - 2 \De_D$, $\om_a = \om_0 - \De_L + \de_a$, and 
$\om_b = \om_0 - \De_L + 2 \De_U + \de_b$. Here again we have Raman 
resonances on the two-photon transitions $\ket{2} \to \ket{1}$ and 
$\ket{2} \to \ket{3}$. The character of the cross-coupling is, however, 
different, since the $E_a$ field is now resonant with the atomic transition 
$\ket{3} \to \ket{5}$, while the $E_b$ field is detuned from the 
$\ket{1} \to \ket{5}$ resonance by the amount $2\De_D$. The Stark shifts
of the states $\ket{1}$ and $\ket{3}$ due to the off-resonant interaction 
with the fields $E_{d_2}$ and $E_{d_1}$ is given by $|\Om_d|^2/3\De_D$ and
$-|\Om_d|^2/\De_D$, respectively, which can be incorporated in the detunings 
$\de_{a,b}$. Only the perpendicular arrangement of the $\sigma_{\pm}$
polarized weak fields and $\pi$ polarized driving fields in a cold atomic gas
(Fig. \ref{system}-Right Inset) is suitable for the cross-absorption scheme, 
since the frequency of the $E_b$ field exactly matches the frequency of the 
atomic transition $\ket{2} \to \ket{4}$ which, in the case of collinear 
geometry, will induce a strong, resonant, unconditional absorption of that 
field. Equations (\ref{ampls}-\ref{alphas}) still apply upon making the 
indicated changes.

In the cross-absorption case, similarly to the case of cross-phase modulation, 
we solve Eqs. (\ref{ampls}) perturbatively in the weak field limit 
($A_2 \simeq 1$). At the Raman resonance for both fields, $\de_a = \de_b = 0$, 
we then obtain for the imaginary part of the polarizabilities
\begin{equation}
\text{Im} (\alpha_{a,b}) \simeq \frac{\alpha_0 |\Om_{b,a}|^2}{|\Om_d|^2}. 
\label{aIm}
\end{equation}
In deriving Eq. (\ref{aIm}), we have assumed that 
$\ga/\De_D > |\Om_{a,b}|^2/|\Om_d|^2$, which is well satisfied for the 
parameters listed above. Thus, if only one of the fields propagates in the 
medium, its absorption is vanishingly small. By contrast, if both fields are 
present, each of them induces a {\it strong absorption} of the other. 
With the experimental parameters given above, the induced absorption depth 
of a single-photon pulse is $\sim 4.3\times 10^{-3}$ cm, i.e., the fields' 
intensities are reduced by a factor of $e$ after they have interacted 
with only $4 300$ atoms (!).


The treatment outlined here has focused on the essential aspects of our 
scheme, yet certain experimentally relevant issues have to be addressed 
briefly. 
{\it (a) Diffraction}: Tightly focused Gaussian beams $E_{a,b}$ would normally 
diffract over the distance 
$\sigma_{a,b}\om_0/2\pi c \simeq 1.3 \times 10^{-4}$ cm, which is much 
smaller than the propagation lengths necessary for achieving the desired 
phase shift and absorption. One can, however, take advantage of the 
long-distance diffraction-free propagation of weak Bessel beams. 
Alternatively, one can use the focusing properties of EIT. 
{\it (b) Adiabaticity}: The adiabatic solution (\ref{alpha_sltn}) of the 
amplitude equations (\ref{ampls}) is justified by the fact that we have 
considered weak (much less that one photon per atom), slowly varying 
($\tau \gg \ga^{-1}$) fields $E_{a,b}$. In the case of intense and/or 
short-duration pulses, however, only the time-dependent treatment of the 
coupled set of Maxwell and density matrix equations will rigorously solve 
the problem. 
{\it (c) Spectral broadening}: Since the phase shift of each pulsed field is 
proportional to the intensity of the other, it will be maximal at the pulse
peak and vanish at the tails. This will produce frequency chirping of the 
pulses and, therefore, their spectral broadening. One has to take care that 
the resulting spectral width does not exceed the transparency window of the 
EIT resonance $\sim |\Om_d|^2/\ga$.
{\it (d) Entanglement}: In the case of cross absorption, our scheme 
(Fig. \ref{system}-Right Inset) can serve as a very sensitive conditional
photon switch \cite{pc}, whose sensitivity is limited by the free-space shot 
noise and the detector efficiency. In the case of cross-phase modulation, a 
phase shift of $\pi$ with negligible absorption should render the two beams 
fully entangled \cite{lukin}. 

To sum up, we have shown that a simple scheme, comprised of a transverse 
static magnetic field and two optical driving fields, can create a new regime 
of {\it symmetric}, extremely efficient nonlinear interaction of two weak 
pulses in atomic vapor, owing to EIT via Zeeman-split levels. The resulting
giantly enhanced cross-absorption and cross-phase modulation may open the 
road to the development of novel Kerr shutters and phase conjugators, as well 
as to quantum information applications \cite{qinf} based on absorptive or 
dispersive two-field entanglement \cite{harris,zhu,harris_d,lukin,pc}.


We acknowledge the support of the EU (ATESIT Network) and the Feinberg 
School (D.P.).


\begin{references}
\bibitem{comment} The absorption-free propagation length $\zeta$ is limited 
\cite{eit} by the ground-state decoherence $\ga_{\text{gr}}$($\ll \ga$)
to $\zeta < v_g/ \ga_{\text{gr}}$.
\bibitem{dip_mom} In the cold atom setup with circularly polarized 
weak fields $E_{a,b}$, all the dipole matrix elements of the 
$F_L=1 \leftrightarrow F_U=1$ transitions have the same absolute value
$|\mu_{ij}|=\mu$, while in the collinear Doppler-free setup with linearly
polarized weak fields, all the dipole matrix elements of the $\De M = \pm 1$
transitions have the absolute value $\mu/\sqrt{2}$.
\end{references}
\end{document}